\documentclass[aps,showpacs,11pt]{revtex4}
\usepackage{dcolumn}
\usepackage{graphicx}
\usepackage{amsmath}
\usepackage{amsfonts}
\usepackage{amssymb}
\usepackage{psfrag}
\usepackage{wrapfig}
\usepackage{subfigure}
\usepackage{makeidx}
\usepackage{bm}
\usepackage{epsf}
\usepackage{hyperref}
\usepackage{cases}
\usepackage{float}
\makeatletter

\newcommand{\Rmnum}[1]{\uppercase\expandafter{\romannumeral #1}}
\makeatother

\begin{document}

\title{Tricritical point and solid/liquid/gas phase transition of higher dimensional AdS black hole in massive gravity}

\author{ Bo Liu$^{1,2,4}$,\ Zhan-Ying Yang$^{1,3}$,\ Rui-Hong Yue$^{4}$\footnote{ Email:rhyue@yzu.edu.cn}}

\affiliation{
$^{1}$School of Physics, Northwest University, Xi'an, 710127, PR China\\
$^{2}$School of Arts and Sciences, Shaanxi University of Science and Technology, Xi'an, 710021, PR China\\
$^{3}$Shaanxi Key Laboratory for Theoretical Physics Frontiers, Xi'an 710127, PR China\\
$^{4}$Center for Gravity and Cosmology, College of physical science and technology, Yangzhou University, Yangzhou, 225009, PR China
}

\date{\today}

\begin{abstract}
  \indent By considering the fifth order term of the interaction
   potential in massive gravity theory,
   we study the $P-V$ critical behaviors of AdS black hole in
   $d \geq 7$ dimensional space-time,
   and find the tricritical point and
   the solid/liquid/gas phase transition
   in addition to the Van der Waals-like phase
   and the reentrant phase transition of the system.
  The critical phenomena of black holes depend crucially on
  the number $n$ of interaction potential terms.

\end{abstract}



\maketitle

\section{Introduction}
\label{1s}
Thermodynamics of black hole revealing the relationship
between the gravity and thermodynamics has always been
one of the hot topics in theoretical physics since the
Hawking radiation was discovered in 1975
\cite{Hawking:1975}.
In recent two decades,  more and more attentions are
paid on the thermodynamics of black hole in anti-de
Sitter (AdS) space.
One of the reason is that the AdS black hole can be
 in thermodynamical stable equilibrium with a positive
 specific heat,
and has Hawking-Page (HP) phase transition
between thermal gas and black hole in AdS space
\cite{Hawking:1983},
which is absent in asymptotically flat or de Sitter space.
Another reason is that the AdS/CFT correspondence
\cite{Gubser:1998,Maldacena:1998,Maldacena:1999,Witten:1998ac},
connecting  gravity theory in bulk with a conformal
field theory on the boundary of AdS space, is regarded as
a powerful tool for understanding the strongly correlated
system by investigating the classical gravity theory.
In this view, HP phase transition could correspond
to the confinement/deconfinement phase transition in gauge
field theory
\cite{Witten:1998pt}.

In1999, it
was discovered
 \cite{Chamblin:1999f,Chamblin:1999s}
 that the first order phase transition
between large black hole and small black hole(LBH/SBH)
in charged Reissner-Nordstr\"{o}m-AdS (RN-AdS) black
displayed in many respects similar to the
liquid-gas phase transition in Van der Waals fluid.
In fact, it was found that the critical exponents
analogous to those of the Van der Waals fluid
do not depend on the number of dimensions of the
RN-AdS black hole
\cite{NIU:2012},
and the identification of these features might be
just a mathematical analogy rather than physical one.

Moreover, according to the Smarr relation of AdS
black hole obtained by the scaling argument
\cite{Sekiwa:2006,Breton:2005},
the cosmological constant $\Lambda$ should be treated
as a variable
rather than a fixed external parameter.
Once the variation of $\Lambda$  is included in the first law,
  $\Lambda$ should be
 identified with thermodynamical pressure in the geometric units
 $G_N=\hbar=c=k=1$ as
 \cite{Kastor:2009}
 \begin{equation}
 \label{eq:pres cosmlg}
  P=-\frac{\Lambda}{8\pi}=\frac{(d-1)(d-2)}{16\pi l^2},
\end{equation}
where $d$ denotes the dimension of space-time and $l$
is the radius of AdS space-time.
In this case,
the black hole mass $M$ should be treated as enthalpy
rather than the internal energy,
and the thermodynamical volume conjugate with the pressure of
black hole can be calculated by using the standard thermodynamic identities
\cite{Parikh:2006pd,Dolan:2011cf,Dolan:2011cs,Dolan:2011pr,Ballik:2013pd}.
So the phase space in this point of view is usually called
as the extended phase space,
different from the earlier one of
black hole.

A physical first-order (LBH/SBH) phase transition
\cite{Kubiznak:2012}
was discovered in the extended phase space of charged AdS black hole,
where the critical behaviors and exponents are precisely
identical with the Van der Waals
liquid-gas system.
Then,
more interesting new critical phenomena
were shown in higher-dimensional rotating AdS black hole.
For example,
in singly spinning rotating case
\cite{Altamirano:2013pd},
there exists a zero-order reentrant phase transition between
intermediate black hole and small black hole along with
LBH/SBH phase transition,
which is a phase transition from large black holes
to small ones and then back to large one again with increasing
the temperature,
a phenomenon also seen
for Born-Infeld black hole
\cite{Gunasekaran:2012hp},
while in multi-spinning case
\cite{Altamirano:2014cg},
the system  has a small/intermediate/large black hole
(S/I/L BH) phase
transition with  one tricritical (or triple critical) point
reminiscent of the "water-like" solid/liquid/gas phase transition.
More discussions for other kinds of AdS black hole can also be
found in
\cite{Hendi:2013pd,Hansen:2017hp,Cai:2013ly,Dutta:2013sj,
Xu:2014xc,Zou:2014zw,Zhao:2014rd,
Altamirano:2014gx,Mo:2014jc,Zou:2014rd,
Xu:2014jc,Xu:2014lb,Dehghani:2014pd,Lee:2014lb,
Zhang:2015jh,Zhang:2015gg,Zhang:2015pd,Hendi:2016pd,
Kubiznak:2017cg,Kuang:2017pd,Miao:2012ah,Zhang:2017rt}.

The Einstein's general relativity (GR) is universally accepted as
 a beautiful and accurate theory describing the force of gravity.
From the perspective of modern particle physics,
GR can be treated as a unique theory of a massless spin-2 graviton
\cite{Weinberg:1965pg,Boulware:1975cg}.
Despite many successes agreement with observations,
GR  might be searched for alternatives due to the open questions,
such as the cosmological constant problem
\cite{weinberg:1989}
that the cosmological constant of astronomical observations is
many orders of magnitude smaller than estimated in modern theories
of elementary particles,
and the origin of acceleration of our universe indicated from
the supernova data
\cite{Riess:1998,Perlmutter:1999},
and so on.
Massive gravity is a straightforward and natural modification
by simply giving a mass to the graviton,
which can date back to 1939 when Fierz and Pauli
\cite{Fierz:1939}
constructed a linear theory of massive gravity.
Whereas,
an elementary problem is that
the massive theory in non-linear level is always plagued with
the Boulware-Deser ghost
\cite{Boulware:1972if,Boulware:1972fr}.
Fortunately,
a ghost-free massive theory  was  proposed
in references
\cite{Rham:2010gd,Rham:2011tl,Rham:2014mg,Hinterbichler:2012},
known as dGRT massive gravity.
A class of charged black holes
\cite{Vegh:2013,Babichev:2014ac}
and their corresponding thermodynamics
\cite{Cai:2015tb}
in asymptotically AdS space-time were investigated in ghost-free massive gravity.
Then other black holes were also studied in massive gravity ,
 such as rotating black hole
 \cite{Babichev:2014rb,Acena:2018is},
Schwarzschild-dS black hole
\cite{Kodama:2014ss}.
The extensions of the solutions in Born-Infeld-massive gravity
and Gauss-Bonnet-Massive Gravity
were also constructed in
\cite{Hendi:2015eb,Hendi:2016jx,Hendi:2018td,Hendi:2018gb}.
Recently,
the holographical and thermodynamical aspects of  black holes in  massive gravity
were also investigated in
\cite{Zeng:2016he,Hendi:2018ha}.
Moreover,
the Van der Waals-like phase  was discovered  in the extended phase space of
the charged AdS black hole in massive gravity
\cite{Xu:2015hc,Hendi:2017vw}.
The reference
\cite{Zou:2017qn}
showed the relationship between this Van der Waals-like phase transition
and the behavior of quasi-normal modes of charged AdS black hole in massive gravity.
Particularly,
the reentrant phase transition with a triple
critical point was discovered in 6-dimensional AdS black hole
in massive gravity
\cite{Zou:2017re}.

In this paper,
we will concentrate on
the  critical behaviors
of $d \geq 7$ dimensional AdS black hole,
and report the finding of the
more richer critical phenomena,
including Van der Waals-like phase transition,
reentrant phase transition,
water-like solid/liquid/gas phase transition,
and triple critical point.
The organization of this paper is as follows.
In Sec.~\ref{2s},
considering the fifth order term of interaction of massive
gravity in $d \geq 7$ dimensional space-time,
 we present the thermodynamics in extended phase space
of higher-dimensional AdS black hole.
In Sec.~\ref{3s},
we study the behaviors of $d \geq 7$ dimensional AdS black hole,
and reveal the richer critical phenomena in the context of $P-V$
criticality and phase diagram.
Finally, a brief discussion is presented in Sec.~\ref{4s}.

\section{Extended phase space thermodynamics of Higher-dimensional AdS black hole in massive gravity }
\label{2s}
Let us start with the  action for  d-dimensional massive gravity
\cite{Rham:2014mg,Hinterbichler:2012}
\begin{equation}\label{eq:action}
  I=\frac{1}{16 \pi} \int d^d x \sqrt{-g} \Big[R +\Lambda +m^2 \sum_{i=1}^{n} c_i \mathcal {U}_i (g,f)\Big],
\end{equation}
where the last term denotes general form of
the interaction potential with graviton mass $m$,
and $n\leq d-2$  the number of dimensionless coupling
coefficients $c_i$. Moreover,
$f$ is a fixed rank-2 symmetric tensor,
 and $\mathcal{U}_i$ are symmetric polynomials
 of the eigenvalues of the ~$d\times d$\ matrix
$K^{\mu}_{\nu}=\sqrt{g^{\mu\alpha}f_{\alpha\nu}}$,
 and  satisfying the following recursion relation
\begin{equation}
 \mathcal{U}_i=-\sum^{i}_{j=1} (-1)^j \frac{(i-1)!}{(i-j)!} [K^j]\mathcal{U}_{i-j}.
 \end{equation}
Obviously, the first few terms can be read as \cite{Rham:2010gd}
\begin{eqnarray}\label{eq:ui}
  \mathcal{U}_1 &=& [K],   \nonumber\\
  \mathcal{U}_2 &=& [K]^2 - [K^2], \nonumber\\
  \mathcal{U}_3 &=& [K]^3 - 3[K][K^2] + 2[K^3],\\
  \mathcal{U}_4 &=& [K]^4 - 6[K^2][K]^2 + 8[K^3][K] + 3[K^2]^2 - 6[K^4],\nonumber\\
  \mathcal{U}_5 &=& [K]^5-10[K^2][K]^3 +15[K][K^2]^2 +20[K]^2[K^3]-20[K^2][K^3]-30[K][K^4]+24[K^5],\nonumber\\
  &&\ldots\nonumber
\end{eqnarray}
where the square brackets denote traces, i.e.~$[K]=K^{\mu}_{\mu}$.

A static black hole solution of d-dimensional space-time is given as
\begin{equation}\label{eq:matr}
  ds^2 = -f(r) dt^2 +f^{-1}(r) dr^2+ r^2 h_{ij} dx^i dx^j,
\end{equation}
in which\ $h_{ij} dx^i dx^j$\ is the line element for an Einstein space with constant curvature\ $(d-2)(d-3)k$,\ and\ $k=1$, $0$, $-1$\ correspond respectively to a spherical,
Ricci flat, and hyperbolic topology subspace.
Considering the following reference metric
\begin{equation}\label{eq:ref metr}
  f_{\mu\nu}=diag(0,0,c_0^2 h_{ij}),
\end{equation}
the interaction potential Eq.~(\ref{eq:ui}) changes into
\begin{eqnarray}\label{eq:spec ui}
  \mathcal{U}_j &=&\Big(\prod_{k=2}^{j+1}d_{k}\Big) c_0^jr^{-j}
\end{eqnarray}
with positive constant $c_0$   and  the notation $d_k=(d-k)$. It is worth to note that the~$c_5 m^2$ term only
appears in the action for~$d\geq 7$, so we just consider the~$d\geq 7$ dimensional black hole and $n=5$ in this paper.

Then, the metric function is calculated as \cite{Cai:2015tb,Zou:2017qn,Xu:2015hc}
\begin{align}\label{eq:metr func}
  f(r)=& k +\frac{16\pi P}{d_1d_2} r^2
       -\frac{16\pi M}{d_2 V_{d-2}r^{d-3}}
       +\frac{c_0 c_1 m^2}{d_2}r
       + c_0^2 c_2 m^2\nonumber\\
       &+\frac{d_3c_0^3 c_3 m^2}{r}
       +\frac{d_3d_4c_0^4 c_4 m^2}{r^2}
       +\frac{d_3d_4d_5c_0^5 c_5 m^2}{r^3},
\end{align}
here ~$V_{d-2}$ is the volume of subspace spanned by coordinates~${x^i}$,
$M$ is the mass of black hole, and
~$P=\frac{d_1d_2}{16\pi l^2}$ is the pressure.

According to the relation~$f( r_h)=0$ which determines the horizon of black hole, the mass of black hole can be expressed in term of~$r_h$ as
\begin{align}\label{eq:mass}
  M = &\frac{d_2V_{d-2}r_h^{d-3}}{16\pi}\Big[k
        +\frac{16\pi P}{d_1d_2} r_h^2
        +\frac{c_0 c_1 m^2}{d_2}r_h
        + c_0^2 c_2 m^2
        +\frac{d_3c_0^3 c_3 m^2}{r_h} \nonumber\\
       &+\frac{d_3d_4c_0^4 c_4 m^2}{r_h^2}
       +\frac{d_3d_4d_5c_0^5 c_5 m^2}{r_h^3}\Big].
\end{align}
And the Hawking temperature~$T$ and the entropy~$S$ of black hole can be obtained as
\begin{align}\label{eq:temp}
  T=&\frac{1}{4\pi r_h}\Big[d_3 k
     + \frac{16\pi P}{d_2} r^2_h
     +c_0 c_1 m^2 r_h
     +d_3 c_0^2 c_2 m^2
     +\frac{d_3d_4c_0^3 c_3 m^2}{r_h} \nonumber\\
    &+\frac{d_3d_4d_5c_0^4 c_4 m^2}{r_h^2}
     +\frac{d_3d_4d_5d_6c_0^5 c_5 m^2}{r_h^3}\Big]\nonumber\\
  S=&\frac{V_{d-2}}{4} r_h^{d-2}.
\end{align}

Due to the mass of black hole corresponding to the enthalpy of an AdS gravitational system, we can get  the Smarr relation as follow using the scaling method
\begin{eqnarray}\label{eq:Smarr}
  (d-3)M&=&\displaystyle (d-2)TS-2PV
         - \frac{c_0 c_1 m^2 V_{d-2} r_h^{d-2}}{16\pi}
         +\frac{d_2 d_3 c_0^3 c_3 m^2 V_{d-2}r_h^{d-4}}{16\pi}\nonumber\\
         & &\displaystyle +\frac{d_2 d_3 d_4 c_0^4 c_4 m^2 V_{d-2}r_h^{d-5}}{8\pi}
           +\frac{3 d_2 d_3 d_4 d_5 c_0^5 c_5 m^2 V_{d-2}r_h^{d-6}}{16\pi},
\end{eqnarray}
where~$V=\frac{V_{d-2}}{d-1}r_h^{d-1}$ denotes the thermodynamic volume conjugate with the pressure~$P$ in extended phase space.
Moreover, the first law of black hole thermodynamics can be written as the following differential relation
\begin{eqnarray}
  dM&=& \displaystyle
           TdS + VdP
           +\frac{c_0 m^2 V_{d-2}r_h^{d-2}}{16\pi} dc_1
           +\frac{d_2 c_0^2 m^2 V_{d-2} r_h^{d-3}}{16\pi} dc_2
           +\frac{d_2 d_3 c_0^3 m^2 V_{d-2} r_h^{d-4}}{16\pi} dc_3\nonumber\\
      &    &\displaystyle
      +\frac{d_2 d_3 d_4 c_0^4 m^2 V_{d-2} r_h^{d-5}}{16\pi}  dc_4
      +\frac{d_2 d_3 d_4 d_5 c_0^5 m^2 V_{d-2}r_h^{d-6}}{16\pi} dc_5.
\end{eqnarray}

Obeying the thermodynamical formulas,
the Gibbs free energy can be written as
\begin{eqnarray}\label{eq: Gibbs fr en}
  G=M-TS=-\frac{V_{d-2} r_h^{d-3}}{2}
          \left[ \frac{2P r_h^2}{d_1 d_2}
          +\omega_2
          +\frac{2 d_3 \omega_3}{r_h}+\frac{3 d_3 d_4 \omega_4}{r_h^2}
          +\frac{4 d_3 d_4 d_5 \omega_5}{r_h^3}\right].
\end{eqnarray}

\section{Critical Behaviors of Higher-Dimensional Black Hole }
\label{3s}
\subsection{Equation of state}
\label{3s:1}
With the help of Eq.~(\ref{eq:temp}), the equation of state of black hole~$P(V,T)$\ can be written as
\begin{align}\label{eq:state}
  P=&\frac{d_2}{4 r_h }\left[T -\frac{d_3 k}{4\pi r_h}
     -\frac{c_0 c_1 m^2}{4\pi}
     -\frac{d_3 c_0^2 c_2 m^2}{4\pi r_h}
     -\frac{d_3 d_4 c_0^3 c_3 m^2}{4\pi r_h^2} \right.\nonumber\\
    &\left.-\frac{d_3 d_4 d_5 c_0^4 c_4 m^2}{4\pi r_h^3}
     -\frac{d_3 d_4 d_5 d_6 c_0^5 c_5 m^2}{4\pi r_h^4}\right].
\end{align}
According to equation of state[\ref{eq:state}], we can find that the value~$v=\frac{4 r_h }{d-2}\propto r_h$ is regarded as special volume comparing to the Van der Waals fluid equation. For the further convenience, we introduce the following denotations
\begin{eqnarray}\label{eq:denotation}
  \hat{T} &=& T-\frac{c_0 c_1 m^2}{4\pi};\ \
  \omega_2 = -\frac{k+c_0^2 c_2 m^2}{8\pi};\nonumber\\
  \omega_3 &=& -\frac{c_0^3 c_3 m^2}{8\pi };\ \
  \omega_4 = -\frac{c_0^4 c_4 m^2}{8\pi};\ \
  \omega_5= -\frac{c_0^5 c_5 m^2}{8\pi}
\end{eqnarray}
where~$\hat{T}$ is called shifted temperature and could be negative value.

Using the condition of inflection point in Van der Walls system
\begin{equation}
  \frac{\partial P}{\partial r_h}\left|_{\hat{T}=\hat{T_c},r_h=r_c}\right.
  =\frac{\partial^2 P}{\partial r_h^2}\left|_{\hat{T}=\hat{T_c},r_h=r_c}\right.
  =0,
\end{equation}
we can receive the critical shifted temperature and the equation of
critical radius of black hole as
\begin{align}
  \hat{T_c}=-\frac{2 d_3 }{r_c}\Big[2\omega_2
  +\frac{3 d_4 \omega_3 }{r_c}
  +\frac{4 d_4 d_5 \omega_4 }{r_c^2}
  +\frac{5 d_4 d_5 d_6 \omega_5 }{r_c^3}\Big]  \label{eq:cri temp}\\
  \omega_2 r_c^3
  +3 d_4 \omega_3 r_c^2
  + 6 d_4 d_5 \omega_4 r_c
  +10 d_4 d_5 d_6 \omega_5 =0\label{eq:criradi}
\end{align}

In the case of $\omega_2=0$, or $\omega_5=0$ ,
one can easily find
that the Eq.(\ref{eq:criradi}) has at most two positive real roots
corresponding to the critical radii of black hole,
 the critical behaviors are similar to the ones in reference\cite{Zou:2017re}. We neglected it and  focus only on the case of $\omega_2\neq0$, $\omega_5\neq 0$ ,
where exists probably three critical radii.

For the simplicity, denoting
\begin{equation}\label{eq:params}
\alpha=3d_4 \omega_3/\omega_2,\ \
\beta=6 d_4 d_5 \omega_4/\omega_2,\ \
\gamma=10 d_4 d_5 d_6 \omega_5/\omega_2,
\end{equation}
the critical temperature and pressure  become into
\begin{eqnarray}
  \hat{T_c}&=&-\frac{2 d_3 \omega_2}{r_c}\Big[2
  +\frac{\alpha }{r_c}
  +\frac{2 \beta }{3 r_c^2}
  +\frac{\gamma }{2 r_c^3}\Big] \nonumber \\
   \hat{P_c}&=&-\frac{2 d_2 d_3 \omega_2}{r_c^2}\Big[2
  +\frac{5\alpha }{3 r_c}
  +\frac{ \beta }{ r_c^2}
  +\frac{4\gamma }{5 r_c^3}\Big] , \label{eq:cri pre1}
\end{eqnarray}
which just depends on the dimension of space-time by a factor $d_2d_3$ ($d_3$)
for given parameters~($\omega_2$,$\alpha$,$\beta$,$\gamma$).

It is easy to find that the Eq.(\ref{eq:criradi}) has at most two positive roots in the case $\alpha\geq 0$.
On the other hand, there would exist three real toots if $\alpha<0 $. Since  the  value of $\alpha$ could be set to be one by re-scaling $r_c$,  we will assume  $\alpha=-1 $
in following discussion.  Based on the requirement of three real roots of the Eq.(\ref{eq:criradi}), we can obtain the relation between the parameters and the number of positive real toots
in TABLE.\ref{tab:num pts}.

\begin{table}[!htb]
\centering
\begin{tabular}{|c|c|c|c|c|c|}
\hline
parameters&
\multicolumn{5}{|c|}{$\alpha=-1$}\\
\hline
$\beta$&
$1/4<\beta<1/3$&
\multicolumn{2}{|c|}{$0<\beta\leq1/4$}&
\multicolumn{2}{|c|}{$\beta\leq0$}\\ 
\hline
$\gamma$&
$\gamma_-<\gamma<\gamma_+$&
$\gamma_-<\gamma<0$&
$0<\gamma<\gamma_+$&
$0<\gamma<\gamma_+$&
$\gamma_-<\gamma<0$\\
\hline
number of $r_c$& 3& 3& 2& 2& 1\\
\hline
\end{tabular}
\caption{The behaviors of the critical radii
   for different values of $\beta$ and $\gamma$  with
     $\gamma_{\pm}=
   \frac{2-9\beta \pm 2(1-3\beta)^{3/2}}{27}$
   }
\label{tab:num pts}
\end{table}

Due to the positive pressures,
when all of the pressures $P_{c1,2,3}$ corresponding to the critical radii above
 are positive,
we can obtain three critical points in the P-V process,
and when two of $P_{c1,2,3}$ are positive,
we will get two critical points, and so on.
For example, taking d=7 and letting $\omega_2=\pm1$
we will get  three positive corresponding pressures while
$\omega_2=-1$,\ \
$2/9<\beta<1/4$,\ \
$\gamma_p<\gamma<0$, or
$\omega_2=-1$,\ \
$1/4<\beta<1/3$,\ \
$\gamma_p<\gamma<0$,\ \
where $\gamma_p$ must be determined numerically.

According to the Gibbs free energy Eq.(\ref{eq: Gibbs fr en}) and the equation of state Eq.(\ref{eq:state}), let us now  study the possible phase transitions
of this system.

\subsection{Van der Waals-like phase transition}
\label{3s:2}

\begin{figure}[H]
  \subfigure[$P-r_h$]{\label{fig:1:1} 
  \includegraphics[width=7.4cm]{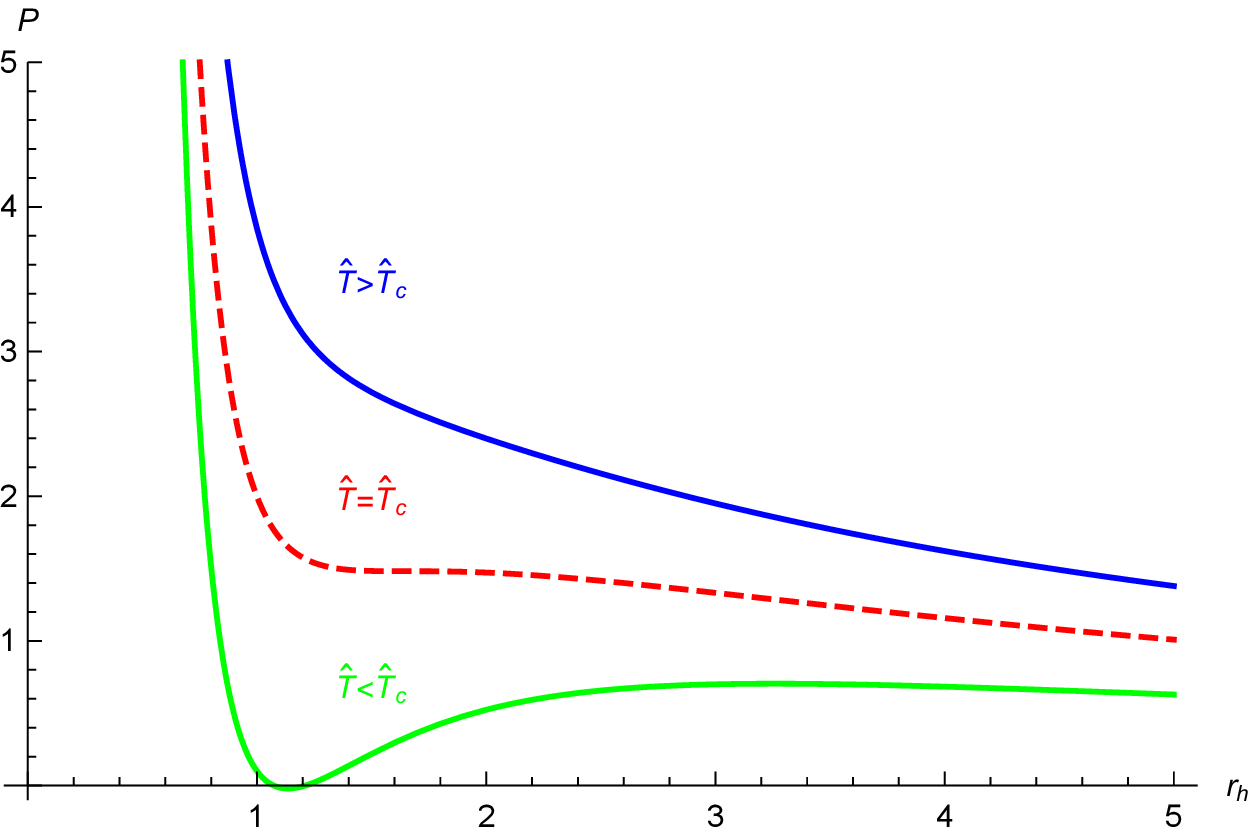}}%
  \hfill%
  \subfigure[$G-\hat{T}$]{\label{fig:1:2} 
  \includegraphics[width=7.4cm]{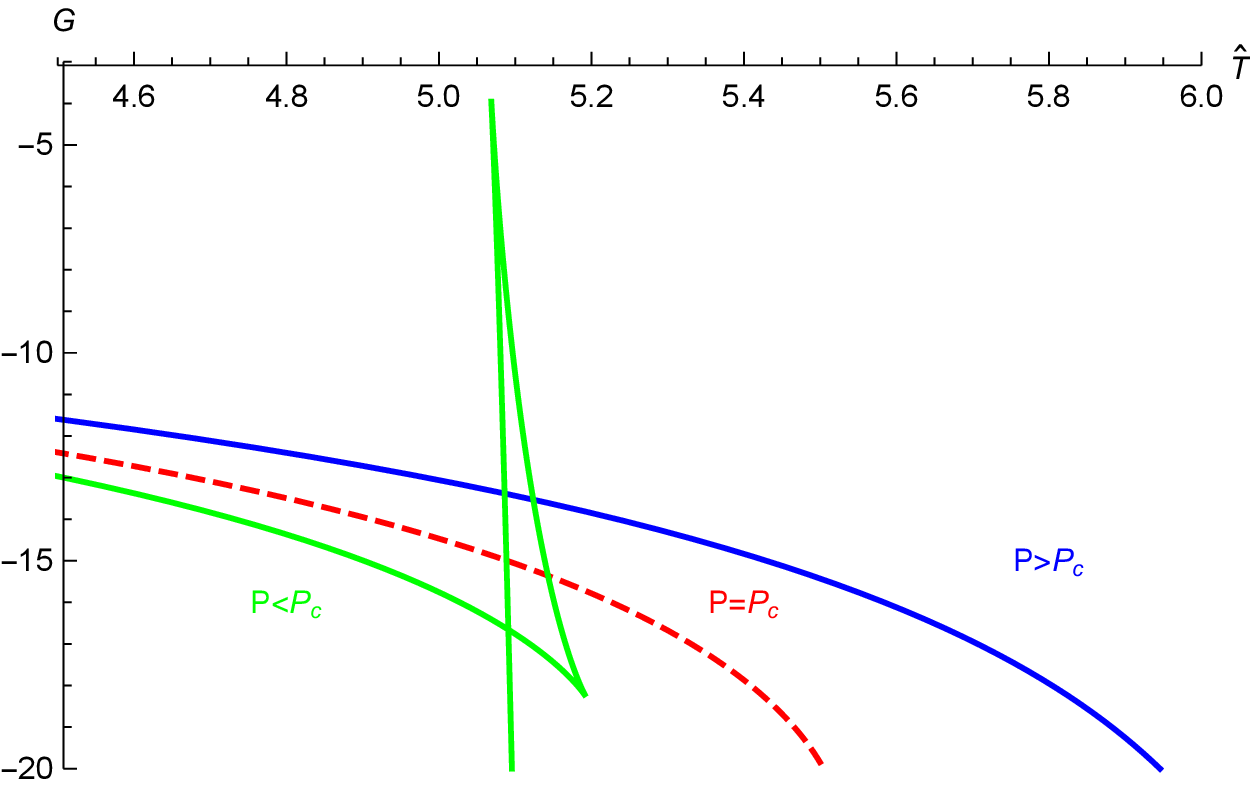}}
  \caption{ The $P-r_h$ and $G-T$ diagrams in d=7
  with $\omega_2=-1$, $\alpha=-1$, $\beta=-1$, $\gamma=-0.1$.
  }
  \label{fig:1}
\end{figure}
In the case of one critical radius of Eq.(\ref{eq:criradi})
in TABLE.\ref{tab:num pts},
the critical behaviors of black hole are analogous to that of the standard Van der Waals-like system as displayed in FIG.{\ref{fig:1}}.

In FIG.\ref{fig:1:1},
there is only one critical isotherm (red dashed line) when $\hat{T}=\hat{T}_c \approx 5.51733$,
and  exits a inflection point in the isotherm (the green solid line) when $\hat{T}<\hat{T}_c$.
Moreover,
in FIG.\ref{fig:1:2} displaying the behaviors of $G$ with a critical isobar
(red dashed line) when $P=P_c \approx 1.48178$,
the isobar corresponding to $P<P_c$ is depicted with a "swallowtail",
which implies a first-order LBH/SBH phase transition.

\subsection{Reentrant phase transition}
\label{3s:3}

There are two ranges of parameters for
two critical radii
in TABLE.\ref{tab:num pts},
both of which  display  the similar thermodynamical processes
with a reentrant phase transition,
analogous to that
in references
\cite{Zou:2017re,Altamirano:2013pd,Gunasekaran:2012hp}.
In one case,
by setting
  $\beta=0.245$
  and
  $\gamma=0.0001$,
the critical behaviors are investigated as follow.

\begin{figure}[htb]
  \subfigure[$P-r_h$ diagram]{\label{fig:2:1} 
  \includegraphics[width=7.4cm]{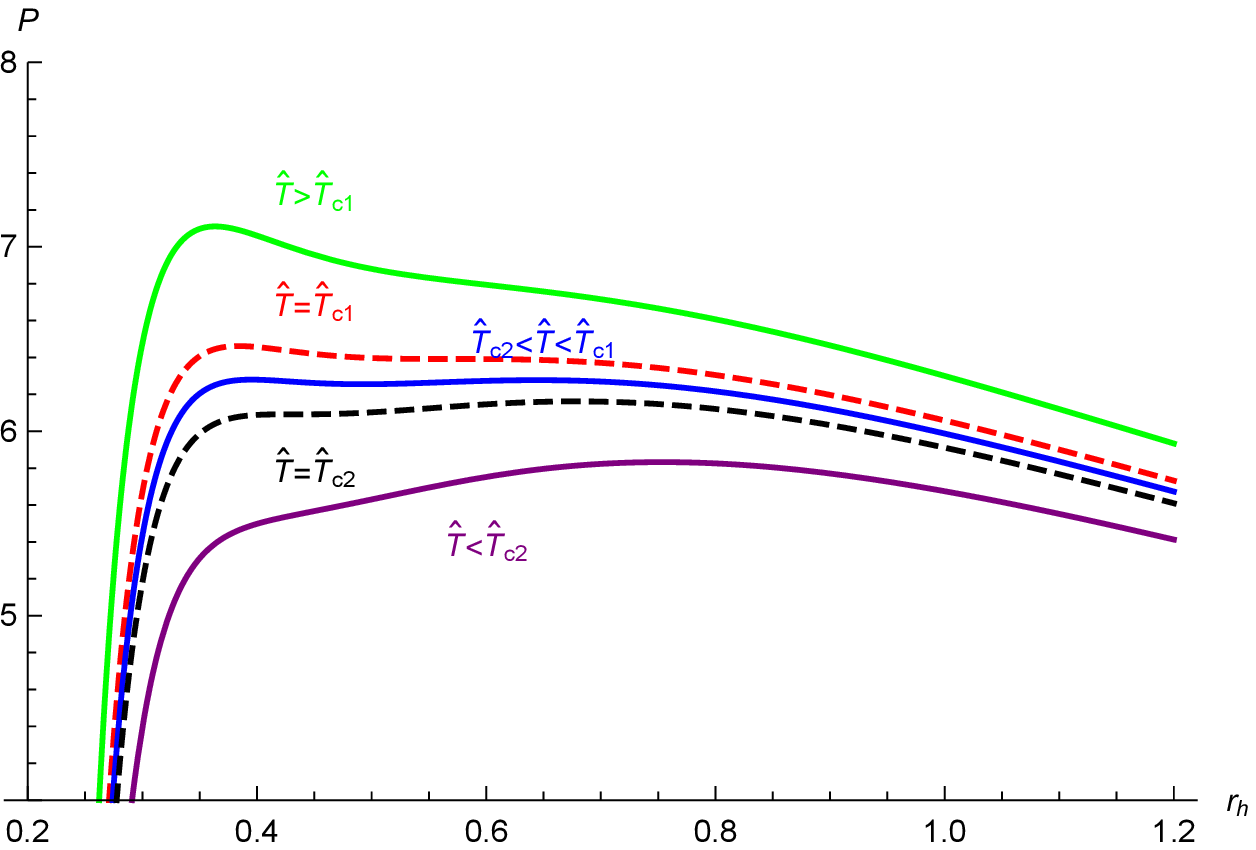}}%
  \hfill%
  \subfigure[$G-\hat{T}$ diagram]{\label{fig:2:2} 
  \includegraphics[width=7.4cm]{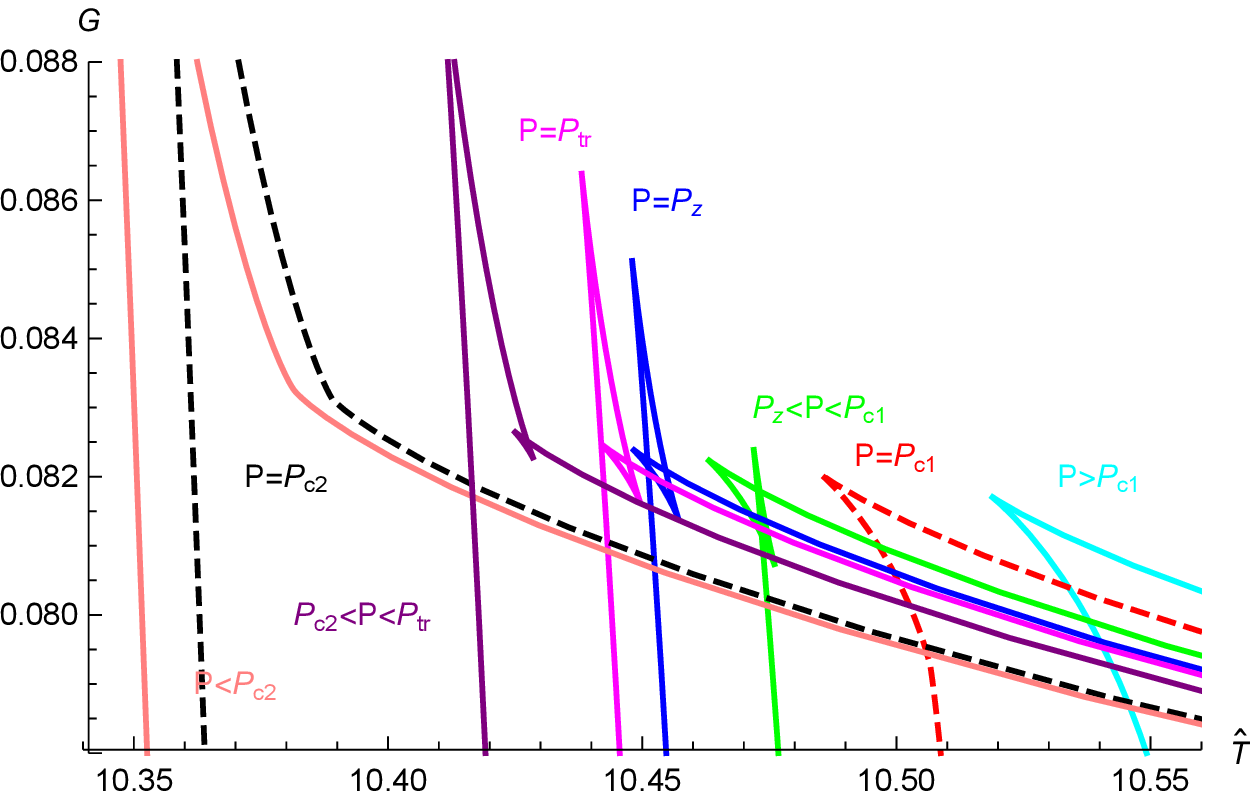}}
  \caption{ The $P-r_h$ and $G-\hat{T}$ diagrams in d=7
  with $\omega_2=-1$,
  $\alpha=-1$,
  $\beta=0.245$,
  $\gamma=0.0001$.
  }
  \label{fig:2}
\end{figure}

In the $P-r_h$ processes indicated in FIG.\ref{fig:2:1},
there exit two critical isotherms,
$\hat{T}=\hat{T}_{c1} \approx 10.50671$
and
$\hat{T}=\hat{T}_{c2} \approx 10.38895$,
corresponding respectively to the red and black dashed lines.
Moreover, there are two inflection points located in each isotherm of the branch corresponding to
$\hat{T}_{c2}<\hat{T}<\hat{T}_{c1}$
as displayed by the blue solid isotherm.

The behaviors of Gibbs free energy $G$
are displayed in FIG.\ref{fig:2:2},
where the radius of black hole increases
from right to left along each isobar,
and
two critical isobars,
the red and black dashed lines,
correspond respectively to the critical pressures
$P=P_{c1} \approx 6.39165$
and
$P=P_{c2} \approx 6.09163$.
The magenta solid isobar corresponds to
the tricritical point,
($\hat{T}_{tr} \approx 10.44195$,
$P_{tr} \approx 6.25396$),
while the blue one to
($\hat{T}_{z} \approx 10.44797$,
$P_{z} \approx 6.27301$).
As the temperature  decreases from right to left,
in the range $P \in (P_{z},P_{c1})$
and
$\hat{T} \in (\hat{T}_{z},\hat{T}_{c1})$,
there is a "swallowtail" indicated by  the green solid isobar signifying a Van der Waals-like phase transition.
In the range $P \in (P_{c2},P_{tr})$
and
$\hat{T} \in (\hat{T}_{c2},\hat{T}_{tr})$,
a "swallowtail" displayed by the purple isobar,
does not correspond to the Van der Waals-like phase transition,
because it is not a "physical" process due to the global minimum of the Gibbs free energy,
so that ($\hat{T}_{c2}$,$P_{c2}$) is not a "really and  physically" critical point.

\begin{figure}[htb]
  \subfigure[Reentrant phase transition]{\label{fig:3:1} 
  \includegraphics[width=7.4cm]{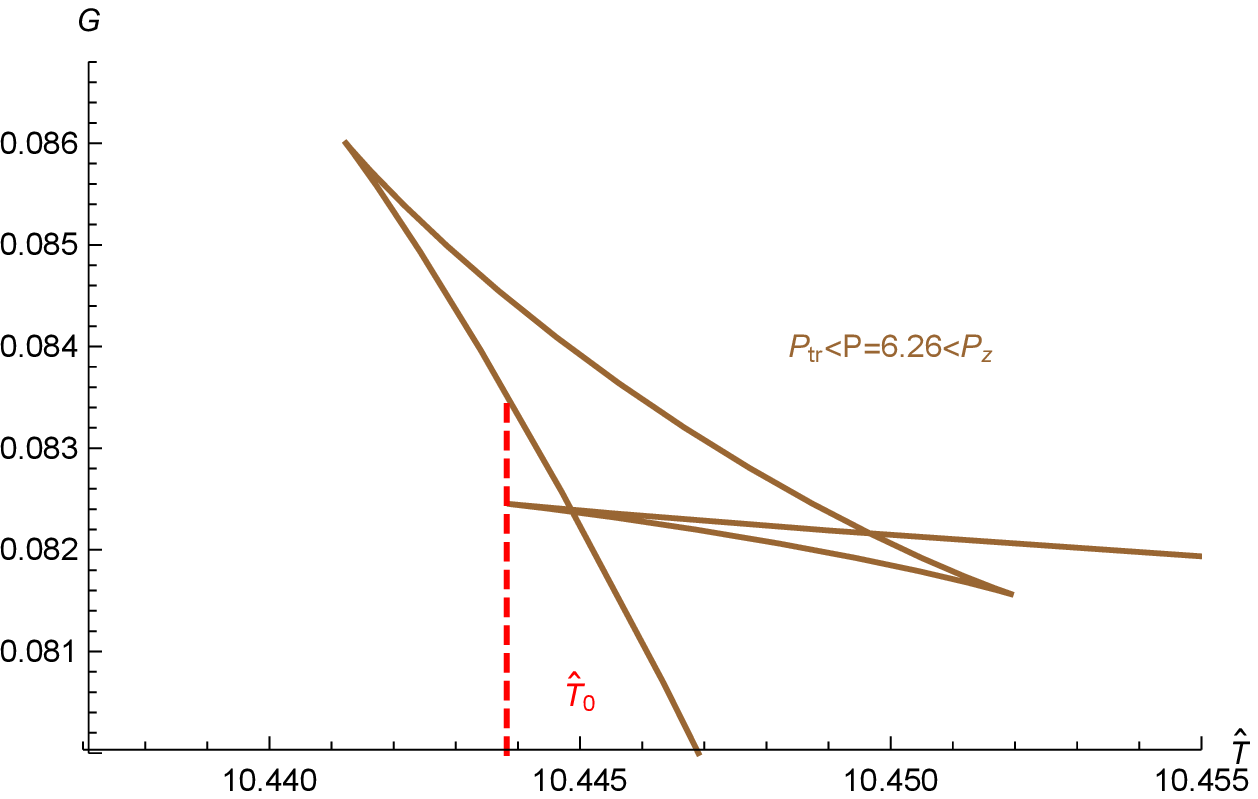}}%
  \hfill%
  \subfigure[$P-\hat{T}$ diagram ]{\label{fig:3:2} 
  \includegraphics[width=8cm]{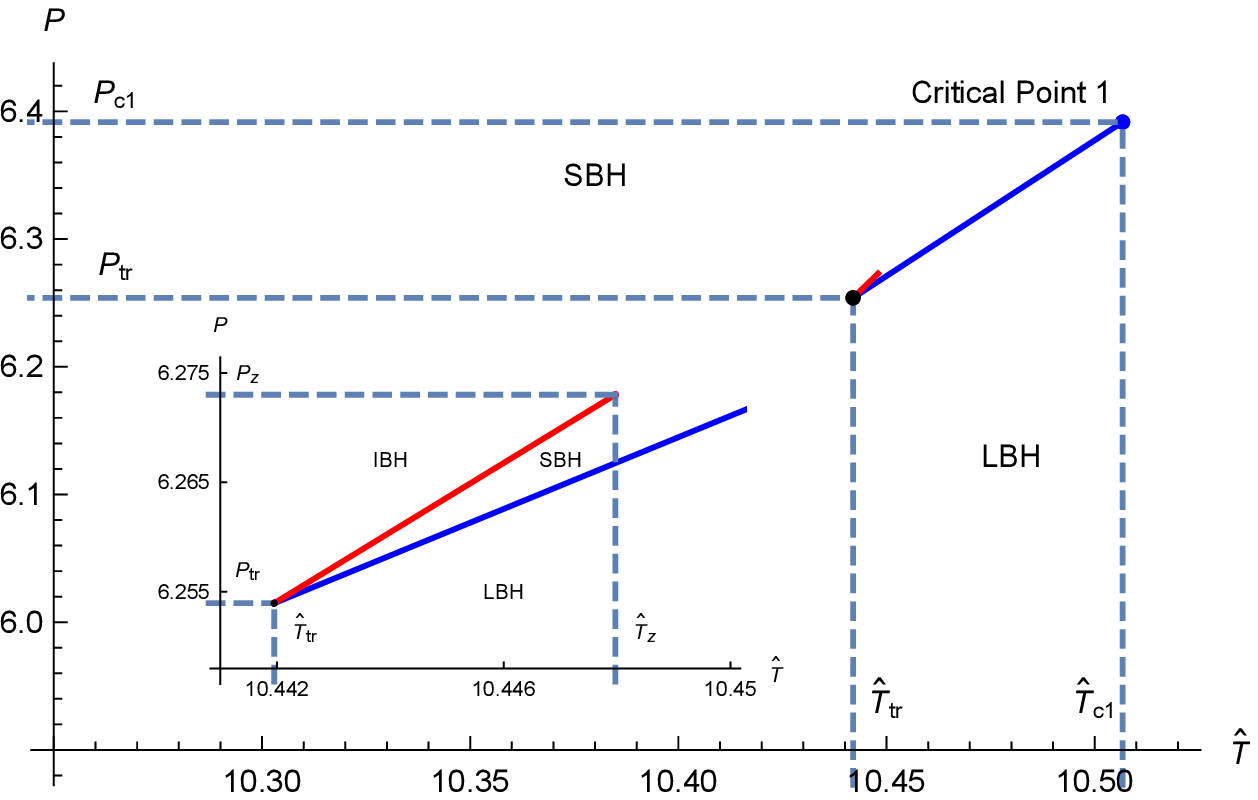}}
  \caption{ The reentrant phase transition for $P=6.26$ and the $P-\hat{T}$ phase diagram in d=7
  with $\omega_2=-1$,
  $\alpha=-1$,
  $\beta=0.245$,
  $\gamma=0.0001$.
  }
   \label{fig:3}
\end{figure}

Especially in the ranges
$P \in (P_{tr},P_{z})$
and
$\hat{T} \in (\hat{T}_{tr},\hat{T}_{z})$,
the global minimum of G is discontinuous as the temperature decreasing.
Take, the isobar $P=6.26\in (P_{tr},P_{z})$
displayed in FIG.\ref{fig:3:1}, for example,
the value of $G$ experiences a finite
jump at $\hat{T}=\hat{T}_0\approx10.44382\in(\hat{T}_{tr},\hat{T}_{z})$
due to the global minimum of the Gibbs
free energy,
which signifies  the zero-order reentrant phase transition between small and intermediate black holes.

It is obviously found in $P-\hat{T}$ phase diagram in FIG.\ref{fig:3:2},
that
the red line in the inset,
initiating from the triple critical point ($\hat{T}_{tr}$,$P_{tr}$)
and terminating at ($\hat{T}_{z}$,$P_{z}$),
corresponds to the coexistence line between IBH and SBH,
while
the blue one,
initiating from ($\hat{T}_{tr}$,$P_{tr}$)
and terminating at ($\hat{T}_{c1}$,$P_{c1}$),
to the coexistence line of SBH and LBH.
 It is worth to state that,
in the range
$P < P_{tr}$
and
$\hat{T} < \hat{T}_{tr}$,
there is no phase transition between SBH (or IBH) and LBH
so that the
($\hat{T}_{tr}$, $P_{tr}$)
is unlike the common one of water.

\begin{figure}[htb]
\center
    \includegraphics[width=8cm]{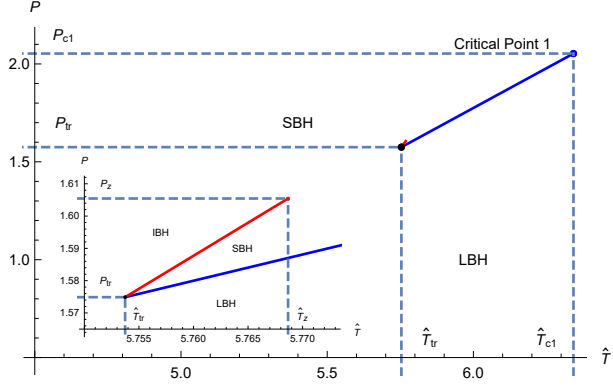}
  \caption{ The $P-\hat{T}$ phase diagram in d=7 with $\omega_2=-1$,
  $\alpha=-1$,
  $\beta=-1$,
  $\gamma=0.85$.
  }
  \label{fig:4}
\end{figure}

In another case,
by setting
  $\beta=-1$
  and
  $\gamma=0.85$,
the critical behaviors are  analogous to those in preceding case
(
  $\beta=0.245$
  and
  $\gamma=0.0001$),
since the $P-\hat{T}$ phase diagram
in FIG.\ref{fig:4} is obviously similar to the one in FIG.\ref{fig:3:2}.
In FIG.\ref{fig:4},
the "physical" critical point is
($\hat{T}_{c1} \approx 6.34216$,$P_{c1} \approx 2.05297$),
the tricritical point is ($\hat{T}_{tr} \approx 5.75374$,
$P_{tr} \approx 1.57488$),
and
the termination point of
reentrant phase transition is
($\hat{T}_{z} \approx 5.76867$,
$P_{z} \approx 1.60544$).
The  blue line corresponds to the coexistence line between SBH and LBH,
 while the red one in the inset to the one between SBH and IBH.

\subsection{Triple critical point and solid/liquid/gas phase transition}
\label{3s:4}

In the case of three critical radii in TABLE.\ref{tab:num pts},
there are also two cases for different parameters,
both of which  display  the analogous thermodynamical processes
as a solid/liquid/gas phase transition with a tricritical
(triple critical) point.
Therefore,
we just focus on one of them by setting $\beta=0.3$,
  $\gamma=-0.027$.

\begin{figure}[htb]
  \includegraphics[width=7.4cm]{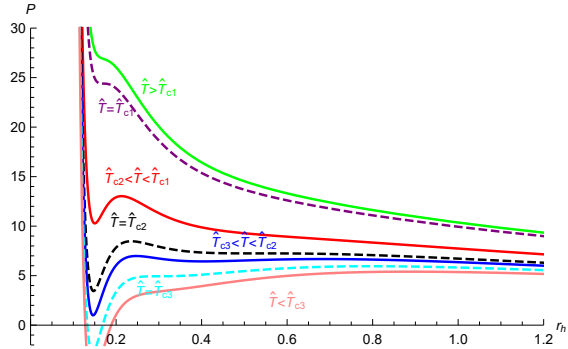}
  \caption{ The $P-r_h$ diagram  in d=7 with
  $\omega_2=-1$,
  $\alpha=-1$,
  $\beta=0.3$,
  $\gamma=-0.027$.
  }
\label{fig:5}
\end{figure}

In FIG.(\ref{fig:5}),
the temperature of isotherm increases from lower left to upper right
in the $P-r_h$ diagram.
Obviously,
the $P-r_h$ diagram, which is more complex
than that of the standard Van der Waals system,
has three critical isotherms,
$\hat{T}=\hat{T}_{c1} \approx 13.65904$,
$\hat{T}=\hat{T}_{c2} \approx 11.08718$,
$\hat{T}=\hat{T}_{c3} \approx 10.37037$,
corresponding respectively to purple, black, cyan dashed lines.
Particularly,
there are three inflection points of each isotherm of the branch
in the range
($\hat{T}_{c3}< \hat{T}<\hat{T}_{c2}$),
as indicated by the blue one,
which maybe suggests the existence of the more complex phase structures.

\begin{figure}[H]
  \subfigure[$G-\hat{T}$]{\label{fig:6:1} 
  \includegraphics[width=7.4cm]{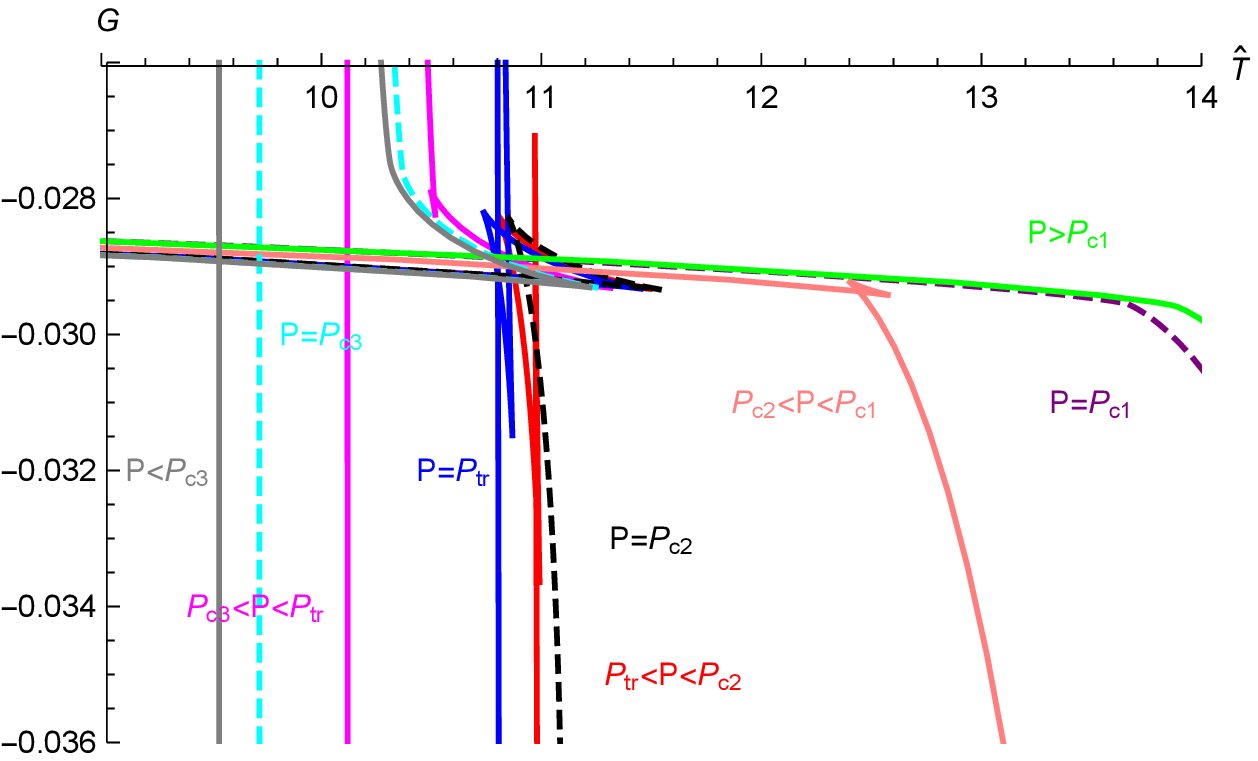}}%
  \hfill%
  \subfigure[Magnification of $G-\hat{T}$ diagram]{\label{fig:6:2} 
  \includegraphics[width=7.4cm]{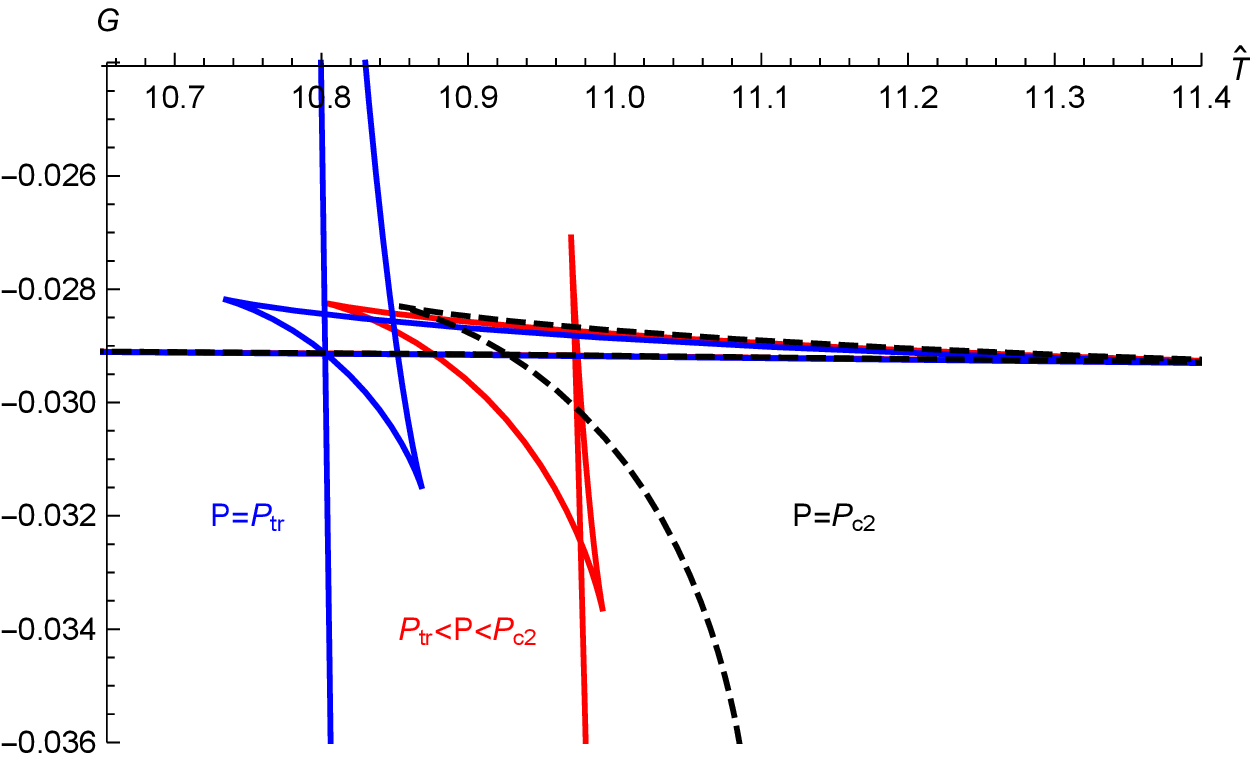}}
  \caption{ The behaviors of $G$ in d=7 with
  $\omega_2=-1$,
  $\alpha=-1$,
  $\beta=0.3$,
  $\gamma=-0.027$.\ \
  A magnification of the blue,
  red, and black isobars is displayed in$(b)$.
  }
\label{fig:6}
\end{figure}

The behaviors of Gibbs free energy $G$ of
black hole are depicted in
Fig.\ref{fig:6},
where the three critical isobars,
$P_{c1} \approx 24.39363$,
$P_{c2} \approx 7.24789$,
$P_{c3} \approx 4.938276$,
are labeled  by the purple, black, and cyan dashed lines respectively.
The blue isobar corresponds to the tricritical point
($P_{tr} \approx 6.64714$,
$\hat{T}_{tr} \approx 10.80263$).
The pressure of isobar increases  from left to right,
and the black hole radius $r_h$ also increases
along each isobar from left to right.
As the temperature decreases from right to left,
for $P>P_{c1}$,
there is no critical behavior as shown
by the green isobar.
For
$P_{c2}<P<P_{c1}$ and
$\hat{T}_{c2}<\hat{T}<\hat{T}_{c1}$,
the pink isobar has one swallowtail, implying a first-order Van
der Waals-like phase transition.
For
 $P_{tr}<P<P_{c2}$
 and $\hat{T}_{tr}<\hat{T}<\hat{T}_{c2}$,
the behaviors are
displayed by the red isobar in FIG.\ref{fig:6:2},
which shows two swallowtails,
corresponding to the coexistence of two first-order phase transitions--SBH/IBH and
IBH/LBH phase transitions.
Until at $P=P_{tr}$ and
$\hat{T}=\hat{T}_{tr}$ in blue isobar
in FIG.\ref{fig:6:2},
the two swallowtails merge with each other,
 corresponding to the tricritical point of the small,
intermediate and
large black holes.
As the temperature continuously decreases,
for
$P_{c3}<P<P_{tr}$
and $\hat{T}_{c3}<\hat{T}<\hat{T}_{tr}$,
the magenta isobar describing the behaviors is also with two
 swallowtails,
but the "small" swallowtail on the upper-right side
is not a "really and physical" critical
process because of the globally minimization of the Gibbs free energy $G$,
  similar to the situation of the reentrant phase transition
  in Subsec. \ref{3s:3}.
So only the "big" one left corresponds to the
first-order phase transition.
Finally,
there is also only one swallowtail occurring
in the case $P<P_{tr}$,
so that the critical point ($P_{c3}$, $\hat{T}_{c3}$)
is not a physically critical one.
Therefore,
there exist
two physically critical points
($P_{c1}$,$\hat{T}_{c1}$) and
($P_{c2}$,$\hat{T}_{c2}$) and one tricritical point ($P_{tr},\hat{T}_{tr}$).

\begin{figure}[H]
  \subfigure[$P-\hat{T}$ diagram]{\label{fig:pt:1} 
  \includegraphics{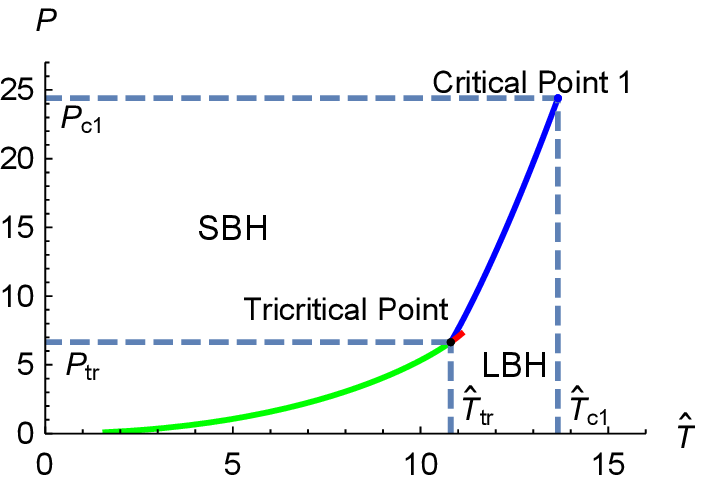}}%
  \hfill%
  \subfigure[Magnification of $P-\hat{T}$ diagram]{\label{fig:pt:2} 
  \includegraphics{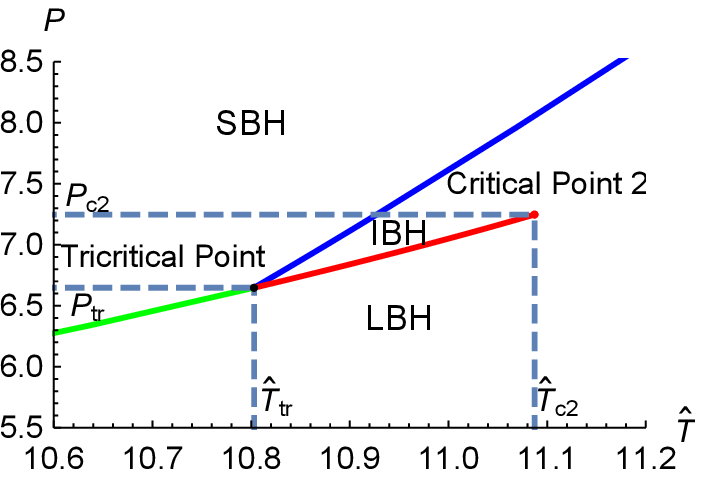}}
  \caption{ The $P-\hat{T}$ diagrams in d=7
  with $\omega_2=-1$,
  $\alpha=-1$,
  $\beta=0.3$,
  $\gamma=-0.027$.}
   \label{fig:pt}
\end{figure}

This tricritical behavior of S/I/L BH phase transition is obviously depicted by $P-\hat{T}$ phase diagrams in Fig.\ref{fig:pt}.
These diagrams are analogous to that of the
water-like solid/liquid/gas phase transition with a triple critical point.
The SBH/LBH coexistence line,
denoted in green color
and terminating at the tricritical point
($\hat{T}_{tr}$,
$P_{tr}$),
corresponds to the solid/gas coexistence line of water.
The LBH/IBH and IBH/SBH coexistence lines,
depicted respectively by the red and blue ones,
do not extend to infinity
and terminate respectively at the critical points 1
($\hat{T}_{c1}$,
$P_{c1}$)
and 2
($\hat{T}_{c2}$,
$P_{c2}$),
corresponding respectively to the solid/liquid   and liquid/gas coexistence ones of water.
Moreover,
the join point (or the tricritical point) of the the three coexistence lines
corresponds to the state of coexistence of
small/intermediate/large black holes with
a special critical value of temperature and pressure.
It is necessary to note that this triple critical  point is a more "common"
one than that in reentrant phase transition,
because there still exists a coexistence line
between SBH and LBH when
$0<\hat{T}<\hat{T}_{tr}$ and
$0<P_{c1}<P_{tr}$, which is not seen in the reentrant phase transition
in FIG.\ref{fig:3:2}
and FIG.\ref{fig:4}.

\begin{figure}[H]
  \subfigure[$P-r_h$]{\label{fig:7:1} 
  \includegraphics[width=7.4cm]{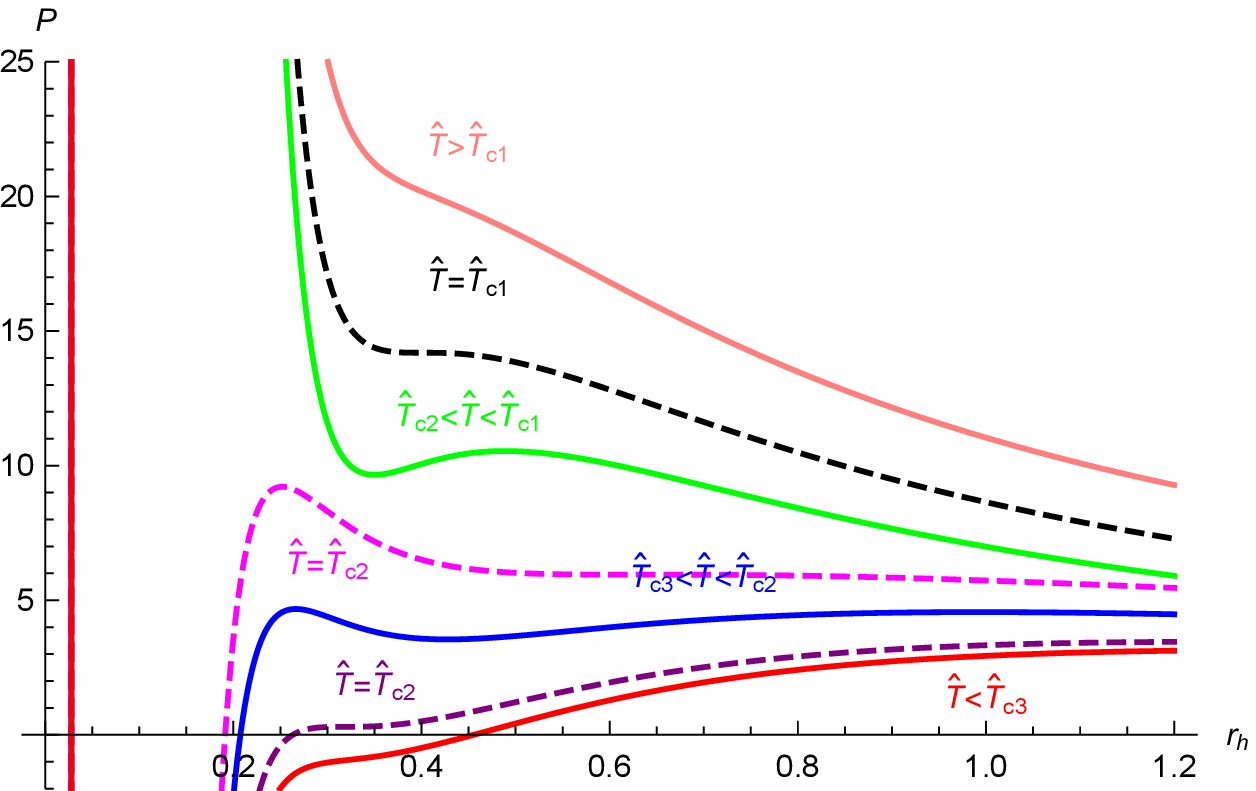}}%
  \hfill%
  \subfigure[$G-\hat{T}$]{\label{fig:7:2} 
  \includegraphics[width=7.4cm]{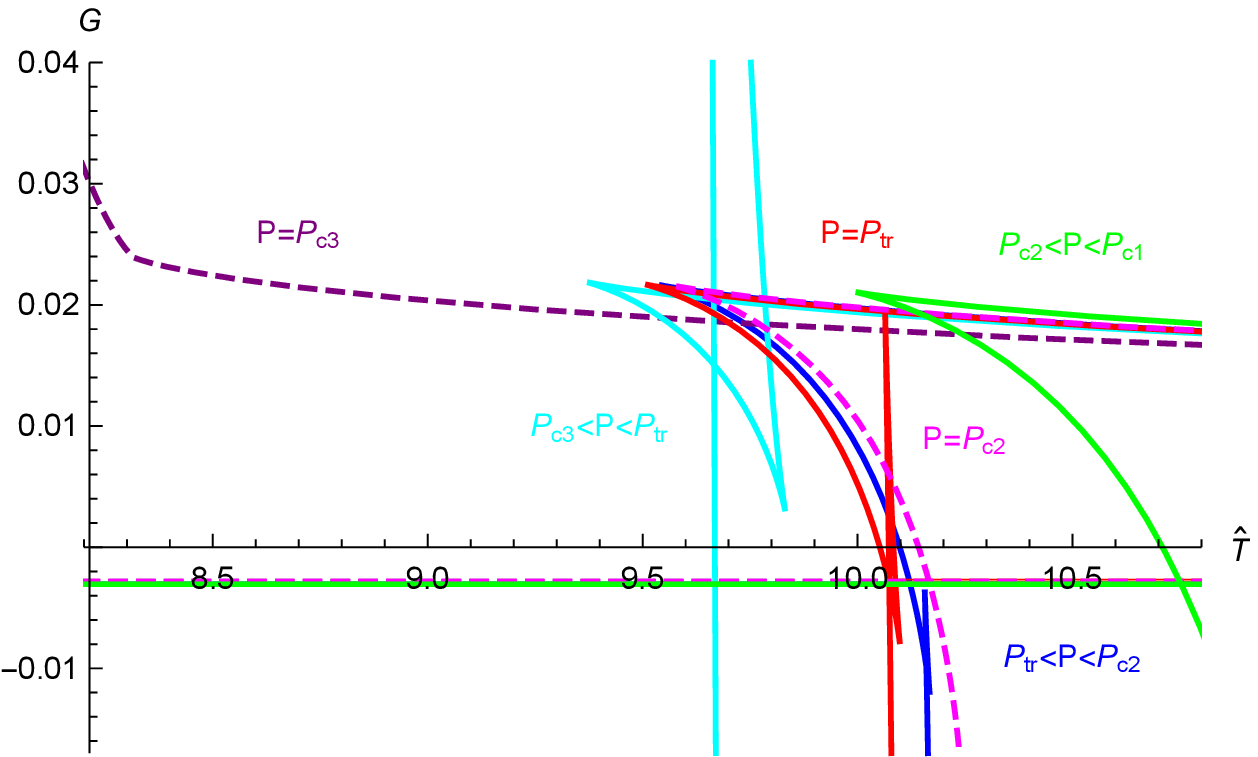}}
  \caption{ The $P-r_h$ and $G-\hat{T}$ diagrams in d=7
  with $\omega_2=-1$,
  $\alpha=-1$,
  $\beta=0.245$,
  $\gamma=-0.009$.}
  \label{fig:7}
\end{figure}

\begin{figure}[H]
  \subfigure[$P-\hat{T}$ diagram]{\label{fig:9:1} 
  \includegraphics{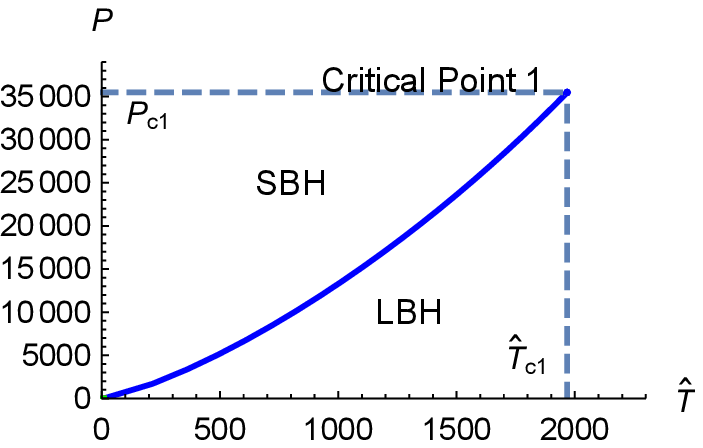}}%
  \hfill%
  \subfigure[Magnification of $P-\hat{T}$ diagram]{\label{fig:9:2} 
  \includegraphics{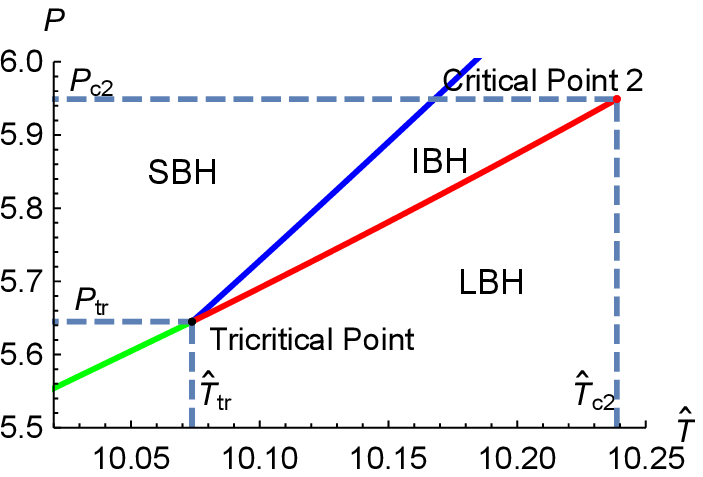}}
  \caption{ The $P-\hat{T}$  diagrams in d=7
  with $\omega_2=-1$,
  $\alpha=-1$,
  $\beta=0.245$,
  $\gamma=-0.009$.}
  \label{fig:9}
\end{figure}

By setting
$\beta=0.245$,
$\gamma=-0.009$
in another case of three critical radii,
the "P-V" processes and critical behaviors of G of BH indicated in FIG.\ref{fig:7} are
respectively similar to those in FIG.\ref{fig:5}
and FIG.\ref{fig:6}).
In FIG.\ref{fig:7:1},
the black,
magenta,
and purple dashed isotherms correspond respectively to the critical temperatures
$\hat{T}_{c1} \approx 1967.34429$,
$\hat{T}_{c2} \approx 10.23876$,
and
$\hat{T}_{c3} \approx 8.31659$.
Because
$\hat{T}_{c1}\gg \hat{T}_{c2},
\hat{T}_{c3}$,
the pressures of the solid pink, dashed black, and solid green lines
are  reduced by 2500 times,
while the radii enlarged by 9 times,
so that all the isotherms can be displayed
in one diagram.
In FIG.\ref{fig:7:2}, the magenta and purple dashed isobars correspond  respectively to
$P_{c2} \approx 5.94913$,
$P_{c3} \approx 0.29619$.
The blue isobar has two swallowtails (big one displayed partially  in Fig.\ref{fig:7:2}),
which implies existence of SBH/IBH and IBH/LBH phase
transitions.
The red isobar denotes to the tricritical point
($T_{tr} \approx 10.07367$,
$P_{tr} \approx 5.64492$ ).
Due to
$P_{c1} \approx 35484.97609\gg P_{c2},
P_{c3}$,
the isobars for $P\geq P_{c1}$ are not displayed in this diagram.
It is more obvious that the magnification of $P-\hat{T}$ phase diagram near tricritical point
in FIG.\ref{fig:9:2} is analogous to the one in FIG.\ref{fig:pt:2}.

\section{Discussion}
\label{4s}
In this paper,
we have shown that, treating the cosmological constant as pressure
and interpreting the corresponding conjugate quantity  as thermodynamic volume,
the higher-dimensional($d\geq 7$) AdS black holes in massive gravity
exhibit  more rich and interesting thermodynamical behaviors,
including Van der Waals-like phase transition,
reentrant phase transition,
solid/liquid/gas phase transition and triple critical point, and so on.
At first,
introducing the fifth term ~$c_5 \mathcal{U}_5$ of interaction potential
in the action of massive gravity theory in the higher dimensional ($d \geq 7$) space-time,
we have obtained a class of solutions of AdS black holes and constructed the thermodynamics of black hole in the extend phase space.
According to the conditions of the inflection point,
we then get the equations of critical temperature and radius basing on the state equation of this system.
Meanwhile, we paid more attention on the case of three real roots of
the equation of the critical radius because of the more complex phase construction,
and obtained the conditions of parameters corresponding to
the number of positive roots or critical radii.
Finally,
we studied the critical behaviors of the Gibbs free energy of black hole.
In the case of only one critical radius,
the critical behaviors of black hole is analogous to those of standard Van der
Waals-like system.
In the case of two critical radii,
it exists a zero-order reentrant phase transition between IBH and SBH due to the
global minimum of $G$.
Particularly in the case of three ones,
there is a water-like solid/liquid/gas phase transition among SBH,
IBH and LBH
with a "common" triple critical point where three phase of black hole can coexists with
the same values of temperature, pressure and the Gibbs free energy.

It is necessary to state that the critical behaviors of AdS black holes
in massive gravity are dependent crucially on the dimension of the space-time through the parameter $n$,
because whether the $c_i$'s term appears in the action of Eq.(\ref{eq:action}) is constrained by $ n\leq d-2$\cite{Hinterbichler:2012}.
For ~$d=5$ ($n\leq 3$) case in the reference\cite{Xu:2015hc}, setting ~$\beta=\gamma=0$ (or $\omega_4=\omega_5=0$ ) due to ~$\mathcal{U}_4=\mathcal{U}_5=0$,
and find that there is at most one critical radius in  Eq.(\ref{eq:criradi}),
thus it has only one-order Van der Waals-like phase transition similar to that in Fig.\ref{fig:1}.
For~$d=6$ ($n\leq 4$) case in reference\cite{Zou:2017re}, corresponding to $\gamma=0$,
the Eq.(\ref{eq:criradi}) has at best two critical radii,
so the zero-order reentrant phase transition occurs analogous to that indicated in FIG.\ref{fig:3} or
FIG.\ref{fig:4}.
Only for $d\geq 7$ could  $\gamma$ not vanish, so that the
 solid/liquid/gase phase transition
 can appear in the case of three critical radii of the $P-r_h$ process.
Specifying to 7 dimensions, in the case of three critical radii,
the solid/liquid/gase phase transition  with a common triple critical
point can occur as displayed in FIG.\ref{fig:pt},
where the S/I/L BH can merge into a single kind with
the same critical value of temperature and pressure.
For $n=5$ and $d>7$ case, the thermodynamical behaviors is similar to ones in $d=7$.
Moreover, the black hole will have more rich thermodynamical structure in high dimensional space-time
since much interaction terms appears in the action, which is valued to be investigated in future.

{\bf Acknowledgements}

We would like to thank Dr. Decheng Zou and Dr. Ming Zhang for many discussions.
This work was supported by the National Natural Science Foundation of China under Grant No.11675139, No.11435006 and No.11875220.

\end{document}